# Injection Scheme with Deflecting Cavity for Ultimate Storage Ring.

J. Kim,[1] G. Jang,[1] M. Yoon,[1] B-H. Oh,[2] J. Lee,[2] J. Ko,[2] Y. Parc,[2] I. Hwang,[2] T. Ha,[2] D. Kim,[2] S. Kim,[3,*] and S. Shin,[2,*]

[1] Department of Physics, POSTECH, Pohang, Gyungbuk 37673, KOREA
[2] Pohang Accelerator Laboratory, POSTECH, Pohang, Gyungbuk 37673, KOREA
[3] FRIB, MSU, Eat Lansing, MI 48824, USA

E-mail: tlssh@postech.ac.kr , kims@frib.msu.edu

**Abstract**. We suggest a new on-axis injection scheme that uses two deflecting cavities to remove the tilt of the stored beam, and to kick the injected beam into the ring. In this new injection scheme, an injected beam with a certain angle is overlaid on the stored beam's position in transverse phase space through the second deflecting cavity. (Note that the injected beam has separated phase with the stored beam in longitudinal phase space). We present theoretical analysis and numerical simulations of the stored beam and injected beam with the new injection scheme.

## 1. Introduction

The ultimate storage ring (USR) based on the multi-bend achromat (MBA) lattice concept may be able to surpass the brightness and coherence that are attained using present third-generation (3G) storage rings. MAX-IV [1] was the first MBA machine; its successor, Sirius [2], is currently under construction. Other projects are being conducted to convert existing 3G machines such as ESRF-II, APSU, Spring-8-II and ALSU [3, 4, 5, 6] to USRs. In these facilities, beam emittance decreases to a few hundred pico-meters, or even to $\leq 100$ pm.

To suppress the dispersions caused by the bending magnets, the USR requires stronger quadrupoles than 3G rings do, so large negative native chromaticities occur in both transverse planes. To correct them, strong chromatic sextupoles are needed. These nonlinear elements will significantly reduce the dynamic apertures, and this is a great challenge for USR lattice design. These nonlinear elements will significantly reduce the dynamic apertures down to ~ 2 mm for many USR lattice designs [4, 5, 6], leading to severe difficulties for injection [7, 8]. Therefore, when the emittance of storage ring is pushed to an extremely small value (i.e., $\leq \sim 100$ pm), dynamic aperture becomes insufficient to enable accumulation-based injection, so only on-axis swap-out injection [9] is workable.

So far, three on-axis injection schemes have been proposed for USR. 1) "Swap-out" injection uses a fast dipole kicker to inject fresh high charge beam onto the closed orbit while the stored beam is extracted. 2) Longitudinal injection with low frequency RF system injects a bunch on-axis at an RF phase that is far away from the synchrotron phase, and with energy slightly higher than the stored beam; the injected bunch damps longitudinally to the synchrotron phase. 3) Longitudinal injection with double-frequency RF system alters two RF voltages to create empty RF buckets that can be taken for on-axis injection; after bunches are injected, the voltage altering process is reversed and the injected bunches can be transferred longitudinally to the main RF buckets. These new injection schemes suggest solutions for beam injection into a small dynamics aperture, but unfortunately the schemes require a tight condition to the injected beam and a kicker that has a very short pulse to avoid disturbing the stored beam.

This paper presents a new on-axis injection scheme for the storage ring with small dynamic aperture. The scheme uses two transverse deflecting cavities (TDCs), and thereby eliminates the need to swap out stored beams and short pulse (a few nanoseconds) kicker. This scheme has the additional benefit of generating a short x-ray pulse between the deflecting cavities. Section 2 introduces the scheme for on-axis injection with two deflecting cavities. Section 3 shows simulation results for the stored beam and injected beam in the lattice of the PAL-USR. Section 4 discusses the result, and Section 5 presents conclusions.

## 2. On-axis injection with transverse deflecting cavity

The longitudinal motion can be described using the longitudinal equation of motion for a relativistic particle

$$\frac{\partial \varphi}{\partial t} = \omega_c \delta + \omega_c \delta^2, \tag{1}$$

$$\frac{\partial \delta}{\partial t} = \frac{1}{E_0 T_0}\left(eV_{rf} - U(\delta)\right). \tag{2}$$

where $\varphi$ is the phase with the given RF frequency, $\delta$ is the relative energy deviation, $E_0$ is the nominal energy, $T_0$ is the revolution period, $V_{rf}$ is the RF voltage and $\omega_{rf}$ is the angular frequency of the ring. $\alpha_c$ and $\alpha_{c2}$ are the 1st order momentum compaction factor and the 2nd order momentum compaction factor, and $U(\delta)$ is the synchrotron radiation energy loss per turn. $U(\delta)$ depends on $\delta$ along turns [10]

$$U = U_0(1 + J_\delta \delta) \tag{3}$$

where

$$D = \frac{\oint \left(\frac{\eta_x}{R^3} + 2\eta_x \frac{K_1}{R}\right) ds}{\oint \frac{1}{R^2} ds} \tag{4}$$

and the fractional part in (4) is calculated from the bending magnet, where $R$ is bending radius, $\eta_x$ is horizontal dispersion function, $K_1$ is the usual focusing parameter [10]. Eq. (1) can be used to calculate the acceptance of longitudinal phase space (Fig. 1, red line). Previous longitudinal injection scheme [11] in injects a beam longitudinally at phase $-\pi$. In this scheme, the injected bunch then longitudinally damps to the synchrotron phase after injection.

Previous longitudinal injection scheme [11] needs (1) a kicker that injects a short pulse (a few nanoseconds) so that stored beam is not disturbed, and (2) tight energy jitter and small energy spread of injected beam so that injected beam is put within the acceptance of the longitudinal phase space at the phase $-\pi$. To overcome these stringent requirements, a new on-axis injection scheme with two transverse deflecting cavities is suggested (Fig. 2). An injected beam with angle is kicked by the second TDC to be put in the same position as the stored beam in transverse phase space. The phase of the TDC is on-crest for injected beam and zero-crossing for the stored beam. The two TDCs work for the stored beam as a system to generate short x-ray pulses that are used to tilt the stored beam by the first TDC and to remove the tilt of stored beam by the second TDC. This scheme does not use a dipole kicker with shorter pulse than the bunch spacing.

RF frequency of TDC is related to the timing offset of the injected beam from the synchrotron phase (Fig. 1). This new on-axis injection scheme must work at $\omega_{TDC} = n \times (\omega_M/2)$, where $\omega_{TDC}$ is the RF frequency of the TDC and $\omega_M$ is the RF frequency of the main cavity. All RF frequencies except 250 MHz for the TDC can be operated with larger energy and timing acceptances for injected beam than in previous longitudinal injection scheme [11] (Fig. 1). Increase in RF frequency reduces RF input power but increases a nonlinear effect that causes leakage on compensation for stored beam. This relationship will be discussed later. Considering available RF power and acceptable leakage on compensation for stored beam, 750 MHz was selected as the RF frequency for the TDC.

The RF parameters of the deflecting cavity (Table 1) were chosen based on the result of a model of 3-cell deflecting cavity with heavy wakefield damping and internal loads [12]. To give a 1.5-mrad transverse kick to an injected beam with 2-μs pulse (Fig. 3), a normal conducting RF cavity with loaded Q ~ 3000 was chosen. Deflecting voltage per cavity is 2 MV and two cavities as the second TDC are used to kick injected beam and to compensate for the tilt on the stored beam. Turn spacing of an injected beam in the PAL-USR is ~ 2 us, so the RF should be pulsed RF with the repetition rate of 1 kHz. Here the reason to choose high repetition rate of 1 kHz rather than general injection repetition rate of a few Hz is for using short pulse X-ray from tilted stored beam at bending beamline. Another two cavities as the first TDC are used to tilt stored beam in upstream section. A feasible option that complies with these specifications is the normal conducting deflecting cavity studied for short pulse X-ray generation at Argonne National Laboratory in collaboration with SLAC National Accelerator Laboratory, the frequency of which is scaled from 2856 MHz to 750 MHz [12]. In this case, the peak RF power required for 2 MV deflecting voltage will be approximately 11 MW with ~ 1 us (FWHM)/ 1 kHz pulses.

## 3. Numerical simulation with PAL-USR lattice

A numerical simulation of the new injection scheme with a deflecting cavity was conducted, In the simulation, the scheme was applied to the lattice of the PAL-USR 3 GeV storage ring. The PAL-USR storage ring is a hybrid 7-bend achromat (H7BA) lattice with a horizontal emittance of 85 pm. The ring with 550 m circumference is composed of 20 symmetrical cells. The relevant parameters are listed in Table 2, and the lattice functions are shown in Fig. 4. The concept of ESRF and APSU lattices was adopted in the PAL-USR lattice. The dispersion was deliberately enlarged between the first and second dipoles, and between the sixth and seventh dipoles; and all chromatic sextupoles were located in this dispersion bump region to reduce the strength required to control the chromaticity. The betatron phase advance between the two dispersion bumps was set to $\Delta\varphi_x \sim 3\pi/2$ in the horizontal plane and $\Delta\varphi_y \sim \pi/2$ in the vertical plane. As a result, non-chromatic effects of the sextupoles are cancelled out naturally. To minimize natural emittance, five-step longitudinal gradient dipoles and reverse bending magnets were considered.

The accelerating voltage in the main RF cavity is 1.2 MV to realize a bucket height of ~5 %. For simplicity, a third harmonic cavity for bunch prolongation is not considered. The stored beam has RMS bunch length = 8 mm, and natural energy spread = 0.1 %. When beam current is 400 mA in multi-bunch operation mode, emittance increase due to intra-beam scattering is 10 % in round-beam mode.

The in-vacuum septum magnet deflects the injected beam by 50 mrad from the beam transport line, and after the septum magnet, the injected beam has 7-mm offset from the stored beam and 1.5-mrad angle. Then the injected beam is assumed to have zero offset and 1.5-mrad angle at the second deflecting cavity and finally, transverse on-axis injection is realized after the second deflecting cavity compensates for the 1.5-mrad injection angle.

To explore the transverse injection dynamics in detail, we used the particle-tracking code ELEGANT [13]. In the tracking, we launched 2,000 particles with the beam distribution from the existing PLS-II linear accelerator (Table 3). Injected bunches with the timing offset were successfully damped to the synchronous phase in the RF bucket (Figure 5). The large transverse emittance (58 nm) of the injected beam was damped toward natural emittance (85 pm) after 60,000 turns.

Emittance degradation of the stored beam was also examined. To investigate those effects, a perfect machine was assumed. For the simulation, we use the extreme case of highest voltage 4 MV for the deflecting cavity (Fig. 3). Of course, the effects diminished as voltage is reduced. Particle distributions in horizontal phase space and horizontal momentum-to-time space (Fig. 6) were calculated with the deflecting cavity on and the deflecting cavity off. In the case of a perfect machine, 30 % emittance degradation occurred.

## 4. Analysis on emittance degradation

This section presents analysis of emittance degradation in the perfect machine, and introduces the effects that arise in a real machine with errors.

The location of TDC2 was fixed in consideration of injection septum and injected orbit, then the position of TDC1 was located to have $n\pi$ phase advance between TDC1 and TDC2 so that I or –I transformation is satisfied. In the PAL-USR lattice, a phase advance of $5\pi$, which corresponds to –I transformation, is set between the two TDCs. However, cancellation was not perfect in either x or x', (Fig. 6).

Analysis of transfer map between TDC1 and TDC2 can explain the imperfect cancellation. With a 3$^{rd}$-order transfer map, beam coordinates at TDC2 can be written as

$$x_i = \sum R_{ij} x_{j0} + \sum T_{ijk} x_{j0} x_{k0} + \sum U_{ijkl} x_{j0} x_{k0} x_{l0} \tag{5}$$

where coordinates with subscript 0 are values at TDC1. First, the linear terms were checked. The value of $R_{22}$ was -0.9457 rad/rad, whereas perfect cancellation requires $R_{22}$ = -1. This deviation of $R_{22}$ from -1 is the source of linearly-correlated residual motion in horizontal phase space; to suppress this motion, the voltage of TDC1 is adjusted to manipulate the momentum coordinate in the initial horizontal phase space. This change improved the cancellation of linearly-correlated residual motion (Figure 7). At TDC1 voltage of 4.23 MV, horizontal emittance after TDC2 had the minimal value 98 pm. The TDCs voltage ratio 4/4.23 (TDC2/TDC1) = 0.9456 is close to the $R_{22}$ ratio 0.9457/1.

Among the nonlinear terms of the transfer map, coefficient $T_{126}$ is dominant. Its value is 67.685 m/rad between TDC1 and TDC2 in the PAL-USR lattice, and diffuses the $x$ coordinate after TDC2 by as much as $T_{126} x_{20} x_{60}$ ( = $T_{126} x'_0 \delta_0$). In the PAL-USR lattice, particles diffusion with $\delta_0 = 0.003$ (edge of 3-σ cut under 0.1 % momentum spread) and $x'_0 = 0.0005$ at TDC1 increase with 0.0001 in position coordinate at TDC2. $x'_0$ is a time-varying kick received by TDC1, so bunch length, voltage of TDC and frequency of TDC also affect diffusion in the position coordinates (Fig. 6a).

The relationship between RF errors and emittance growth, which specifies RF tolerances, is investigated. Voltage difference and beam phase error between TDCs were straightforwardly swept as the source of RF errors. Emittance grew after TDC2 along these errors (Fig. 8). Although emittance growth just after TDC2 seemed negligible ($\pm 2°$ phase errors and $\pm 1$ % voltage error), centroid motion due to the kick by RF error severely affects emittance growth along the ring and turn by turn (Fig. 9): RF error with -1.8° phase deviation and - 0.78-% voltage error causes 200-μm amplitude centroid oscillation and as result emittance growth increases to 3 nm. Beam oscillation due to injection system error in the injection scheme in the present storage ring is ~100 μm ~ 200 μm. Therefore, phase deviation must be maintained within $\pm 2°$ phase errors and $\pm 1$ % voltage error.

## 5. Conclusion

In this paper, a new on-axis top-up injection scheme have been presented for ultimate storage ring. The second transverse deflecting cavity kicks the injected beam toward on-axis in transverse phase space. The first and second transverse deflecting cavities are operated to tilt the stored beam and remove the tilt of the stored beam. Numerical simulation and analysis demonstrated new on-axis injection scheme. In the numerical simulation and analysis, a transverse on-axis injection is realized without causing remarkable disturbance on the stored beam through the ring outside a pair of transverse deflecting cavities. In order to get similar performance with conventional injection system in aspect of injection leakage effect (i.e., 100 ~ 200 um stored beam oscillation), relative phase deviation and voltage error must be operated less than -1.8° and - 0.78 %. Short pulse x-ray, which can be generated from bending magnet between two transverse deflecting cavities, is an additional benefit of new injection scheme.

**Acknowledgments**
We would like to thank M. Aiba (PSI) for providing helpful information and the many useful discussions. This research was supported by POSTECH Basic Science Research Institute. This research was also supported by the Basic Science Research Program through the National Research Foundation of Korea (NRF-2015R1D1A1A01060049).

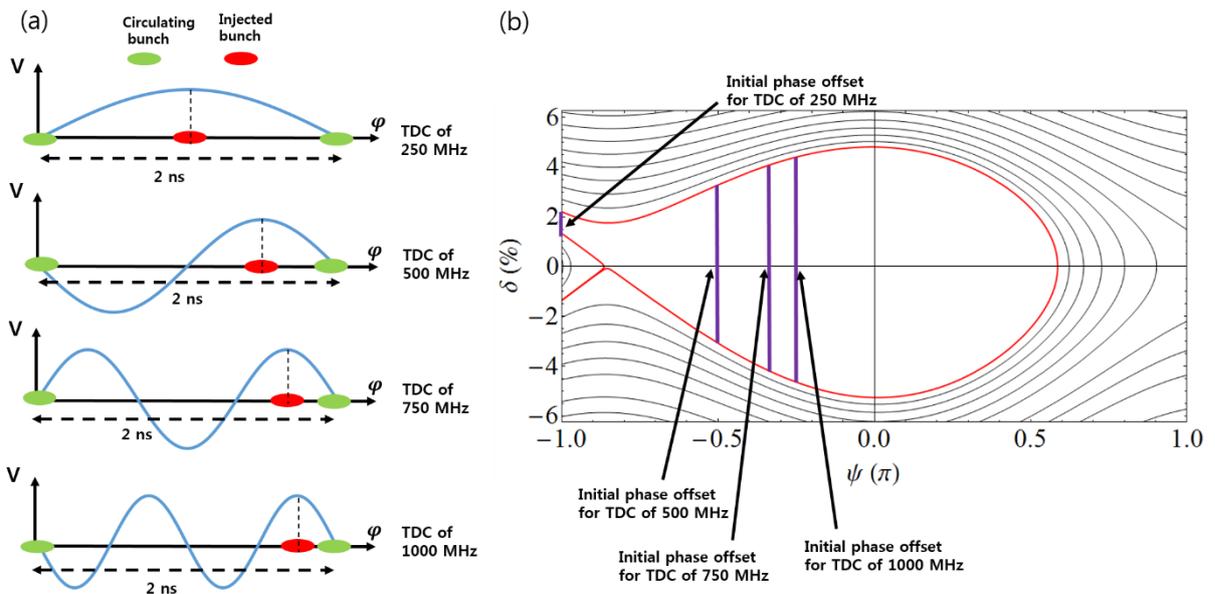

**Figure 1.** Energy acceptance of injected beam along RF frequency of deflecting cavity. Energy acceptance at 750 MHz frequency is already close to energy acceptance for injected beam.

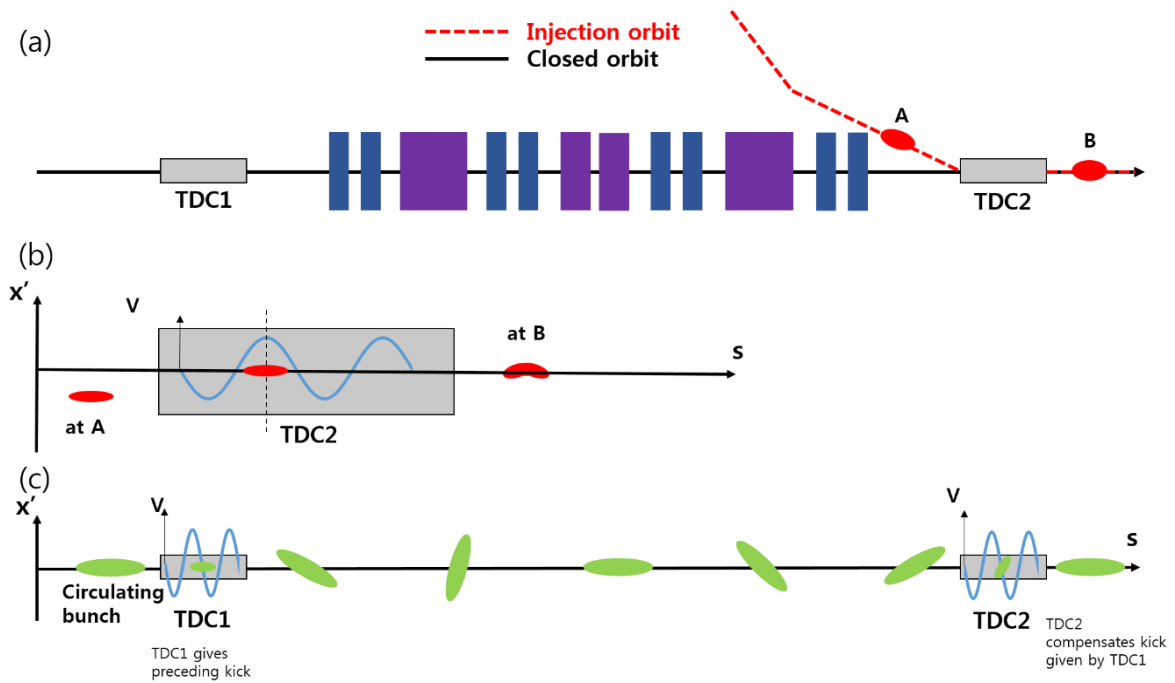

**Figure 2.** Schematics for working principle of on-axis injection with two deflecting cavities.

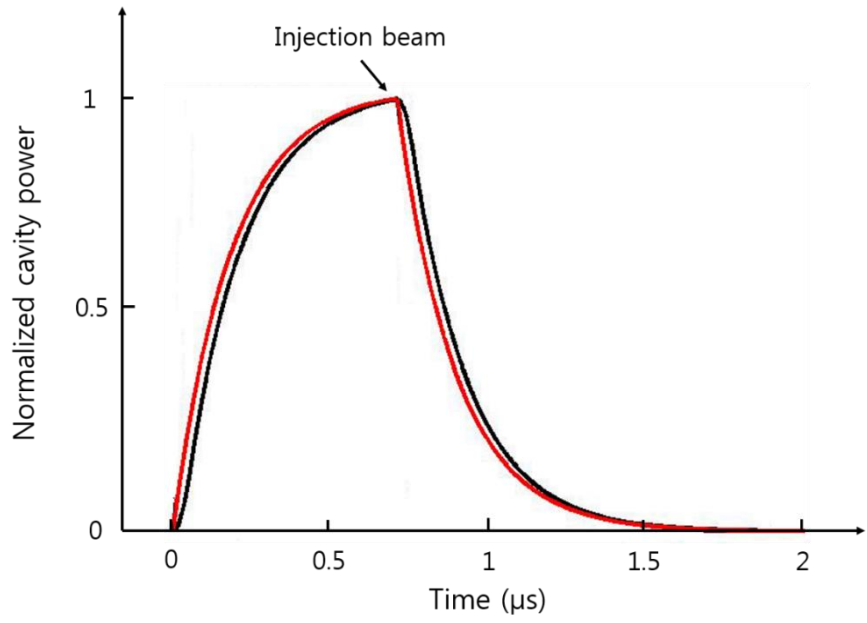

**Figure 3.** Pulse shape of normalized cavity power. Red: ideal square input. Black: 100 ns ramp

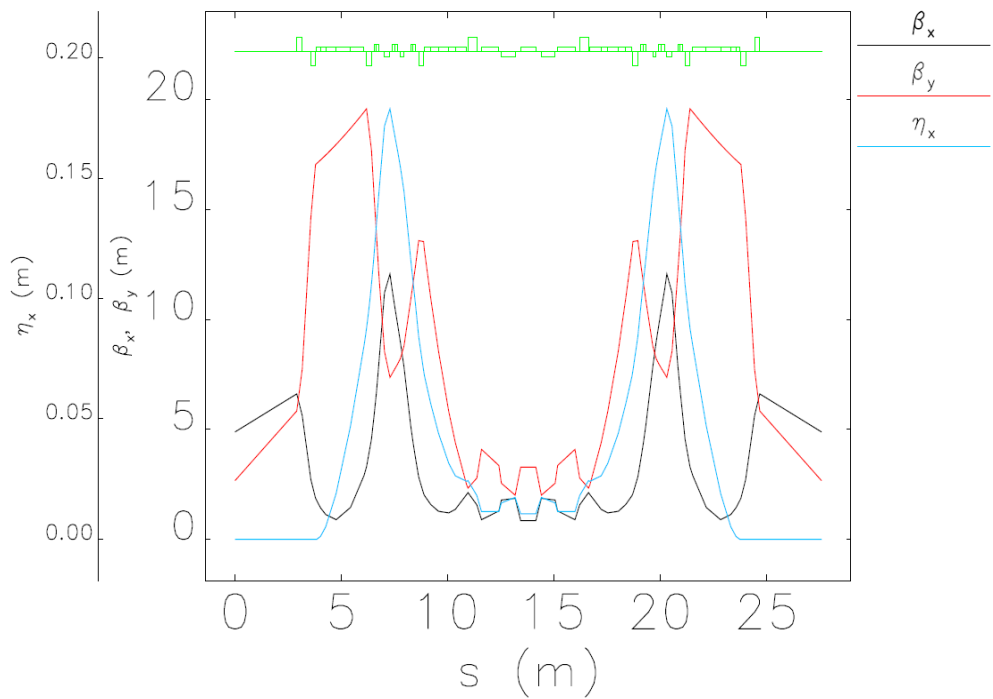

**Figure 4.** Lattice functions for PAL-USR.

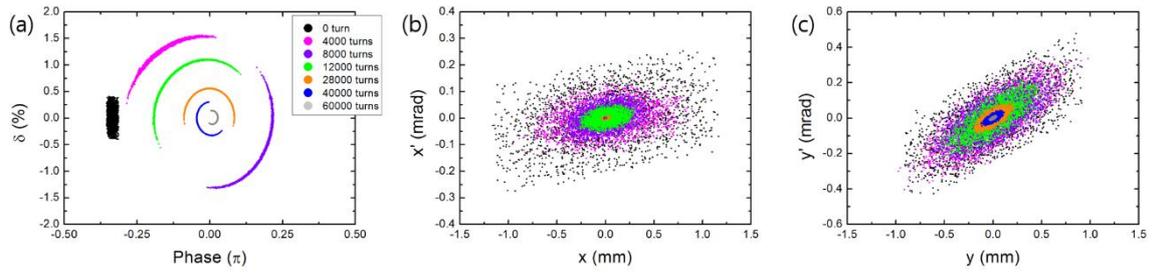

**Figure 5.** Particle distributions in the longitudinal (a), horizontal (b), and vertical (c) planes with up to 60, 000 turns following the injection.

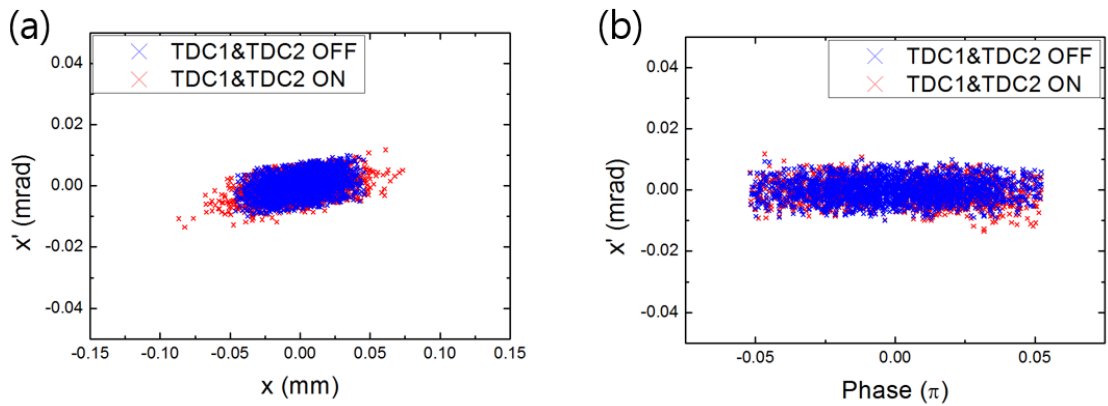

**Figure 6.** Particle distributions in the horizontal phase space (a) and horizontal momentum via phase (b) after the second TDC.

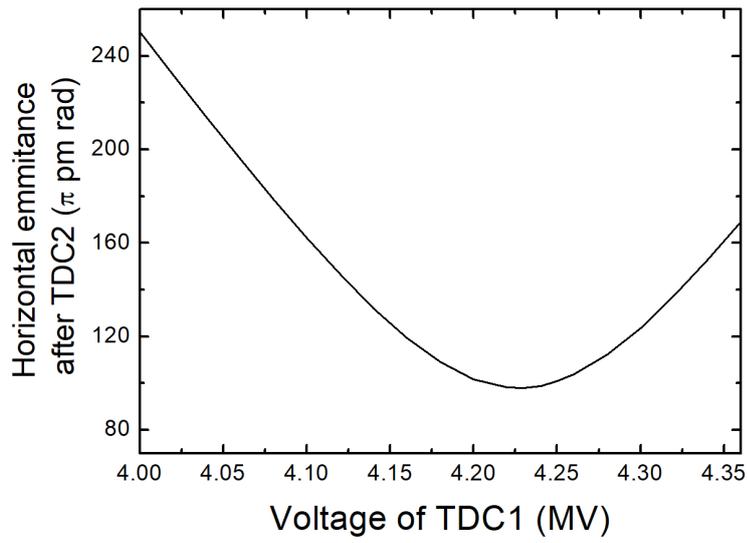

**Figure 7.** Horizontal emittance after TDC2 along voltage of TDC1. Here TDC2 voltage is fixed with 4 MV.

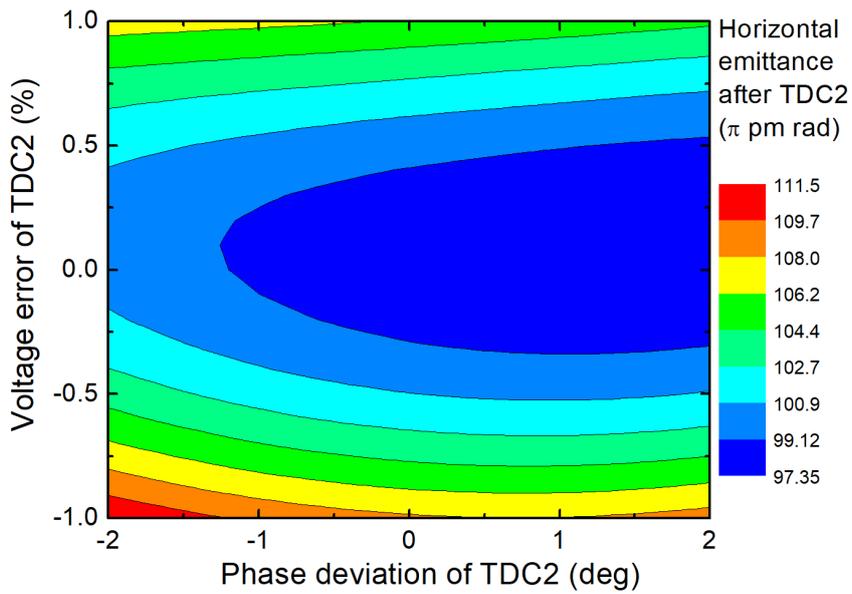

**Figure 8.** Horizontal emittance after TDC2 along relative phase error and voltage error between TDC1 and TDC2.

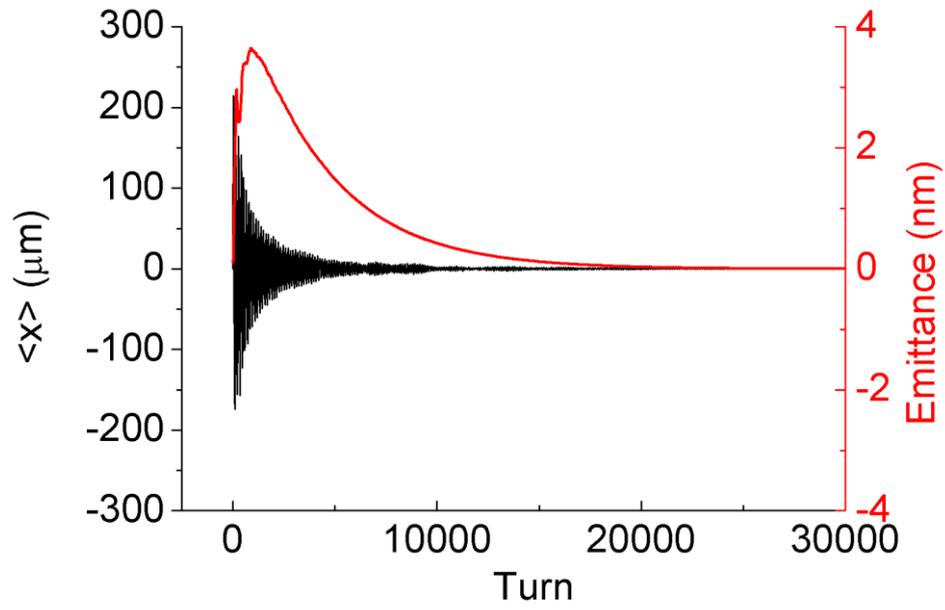

**Figure 9.** Particle distributions in the longitudinal (a), horizontal (b), and vertical (c) planes with up to 60, 000 turns following the injection.

**Table 1.** Deflecting cavity parameters.

| Parameter | Value | Unit |
|---|---|---|
| Frequency | 750 | MHz |
| Deflecting voltage | 2 | MV |
| Cavity type | Normal conducting | - |
| Loaded Q | 3000 | - |
| Pulse duration | 2 | μs |
| Repetition rate | 1 | kHz |

**Table 2.** Parameters relevant to PAL-USR lattice.

| Parameter | Value | Unit |
|---|---|---|
| Energy | 3 | GeV |
| Emittance | 85 / 58 | pm |
| Circumference | 551.8 | m |
| Tune (x/y) | 47.545 / 18.203 | - |
| Natural chromaticity (x/y) | -62.3/-59.5 | - |

**Table 3.** Initial particle distribution.

| Parameter | Value | Unit |
|---|---|---|
| Horizontal emittance | 58 | nm |
| Vertical emittance | 58 | nm |
| Energy spread (rms) | 0.2 | % |
| Bunch length (rms) | 20 | ps |
| Beta function at IP (x/y) | 6.46 / 4.32 | m |